\documentclass[twocolumn]{aastex6}
\usepackage{graphicx}
\usepackage{amsmath}
\usepackage{amssymb}
\usepackage{apjfonts}

\newcommand{\cms}{\,{\rm cm$^{-2}$}\,}
\newcommand{\cmc}{\,{\rm cm$^{-3}$}\,}
\newcommand{\kms}{\,{\rm km\,s$^{-1}$}\,} 
\newcommand{\kmsmpc}{\,{\rm km\,s$^{-1}$\,Mpc$^{-1}$}\,}

\newcommand{\etal}{{ et~al.~}}

\newcommand{\ergs}{\,{\rm erg\,s$^{-1}$}\,}
\newcommand{\ergscm}{\,{\rm erg\,s$^{-1}$\,cm$^{-2}$}\,}
\newcommand{\Ms}{M_\odot}

\newcommand{\Zs}{Z_\odot}

\newcommand{\as}{^{\prime\prime}}

\shorttitle{The Infall of M60 Toward M87} 
\shortauthors{Wood et al.}

\begin{document}

%------------------------------------------TITLE PAGE------------------------

\title{The Infall of the Virgo Elliptical galaxy M60 Toward M87 \\ 
and the gaseous structures produced by Kelvin-Helmholtz Instabilities}

\author{R. A. Wood$^1$, C. Jones$^2$, M. E. Machacek$^2$, W. R. Forman$^2$,
A. Bogdan$^2$, F. Andrade-Santos$^2$, R. P. Kraft$^2$,
A. Paggi$^2$, E. Roediger$^3$  }
\affil{$^1$University of Southampton, Southampton, SO17 1BJ United Kingdom}
\affil{$^2$Harvard-Smithsonian Center for Astrophysics, 60 Garden Street, 
Cambridge, MA 02138 USA}
\affil{$^3$School of Mathematics \& Physical Sciences, University of Hull,
  Hull HU6 7RX, UK\\}
\email{mmachacek@cfa.harvard.edu}

%-------------------------------------------ABSTRACT----------------------------
\begin{abstract}
\noindent We present {\em Chandra} observations of hot gas structures,
characteristic of gas stripping during infall, in the Virgo cluster  
elliptical galaxy M60 (NGC4649) located 1 Mpc east of M87. $0.5-2$\,keV 
{\em Chandra} X-ray images show a sharp leading edge in the surface
brightness $12.4\pm0.1$\,kpc north and west of the galaxy center in 
the direction 
of M87 characteristic of a merger cold front  due to M60's motion through the 
Virgo ICM. We measured a temperature of $1.00 \pm
0.02$\,keV  for abundance $0.5\Zs$ inside the edge
and $1.37^{+0.35}_{-0.19}$\,keV 
for abundance $0.1\Zs$ in the Virgo ICM 
free stream region. 
We find that the observed jump in surface brightness yields  
a density ratio $n_{\rm{in}}/n_{\rm{out}}=6.44^{+1.04}_{-0.67}$
between gas inside the edge and in the cluster free stream region. 
If the edge is a cold front due solely to the infall of M60 
in the direction of M87, we find a pressure ratio of 
$4.7^{+1.7}_{-1.4}$ and  Mach number
$1.7^{+0.3}_{-0.3}$. For $1.37$\,keV Virgo gas, we find a total infall
velocity for M60 of $v_{\rm{M60}}=1030 \pm 180$\kms.  
We calculate the motion in the plane of the sky to be 
$v_{\rm{tran}}=1012^{+183}_{-192}$\kms
implying an inclination angle $\xi=11^{+3}_{-3}$\,degrees.
Surface brightness profiles also show the presence of a faint, 
diffuse gaseous tail. We identify filamentary gaseous wing structures 
caused by the galaxy's motion through the ICM. The structure 
and dimensions of these wings are consistent with simulations of 
Kelvin-Helmholtz instabilities as expected if the gas stripping is 
close to inviscid.
\end{abstract}
%===========================================


%-----------------------------------------CHAPTER1-INTRO-----------------

\section{Introduction}
\label{sec:intro}

The earliest X-ray images taken with {\em Einstein} revealed galaxy
clusters to be far from dynamically relaxed systems (Jones \etal 1979).
Extensive evidence now exists for
subcluster and galaxy infall into clusters.
As galaxies move through the
intracluster medium they undergo hydrodynamical and tidal interactions
of varying severity. 
Surface brightness edges (contact discontinuities in density and
temperature) may be caused by a variety of physical mechanisms. When
seen in cluster galaxies moving through the intra-cluster gas, these  
surface brightness edges  
may be the result of ram-pressure stripping as the galaxy moves through 
the ICM. These `cold fronts' are contact discontinuities at the edge of two gas
regions of different densities and temperatures (e.g. Markevitch \&
Vikhlinin 2007).

From X-ray observations of diffuse galaxy and cluster gas, and the
galaxy redshift, the total galaxy velocity and direction of motion can
be measured, providing one of the few ways that galaxy velocities in the
plane of the sky can be inferred.  The gas densities and temperatures 
are determined from X-ray observations, from which the components of 
thermal pressure on both sides of the contact discontinuity are
derived. Approximating the flow to that of uniform gas about a blunt
body, the total galaxy velocity follows from the 
ratio of thermal pressures at the stagnation point and free stream
region (Vikhlinin \etal 2001).
These cold fronts from ram pressure stripping are often accompanied 
by additional gas features such as wings and stripped gaseous tails
(see e.g. Machacek \etal 2005; 2006; Kraft \etal 2017).
The morphology of these features 
may allow one to constrain the microphysical properties of the
surrounding gas (Roediger \etal 2015a,b).
Cold front edges may also be induced by gas bulk motions (sloshing) caused by
ongoing galaxy mergers. In this case multiple surface brightness
edges or a spiral pattern are often observed, depending on the orientation 
of the merger with respect to the observer's line of sight
(e.g. Markevitch \& Vikhlinen 2007). The measured ratio of 
thermal pressures across a sloshing cold front is
$1$. Finally, surface brightness edges may be due to shocks from
episodic nuclear activity, lifting galaxy gas higher in the galaxy
gravitational potential making it easier to be stripped. Several or
all of these processes may act in concert on galaxies in galaxy
clusters as the galaxies and clusters evolve.
 
The Virgo cluster of galaxies is located $17.2\pm0.6$\,Mpc away
(Mei \etal 2007). 
B{\"o}hringer \etal (1994) show Virgo to
be an evolving cluster; the smooth X-ray distribution towards the
center (M87) and the patchy distribution in the outer regions suggest
the cluster is still forming. The galaxy dynamics of other ellipticals
in Virgo have been studied including NGC4472 (Kraft \etal 2011),
M86 (Randall \etal 2008; Forman \etal 1979), 
and NGC4552 (Machacek \etal 2006; Kraft \etal 2017).

\begin{figure}[htb!]
\begin{center}
\includegraphics[width=3in]{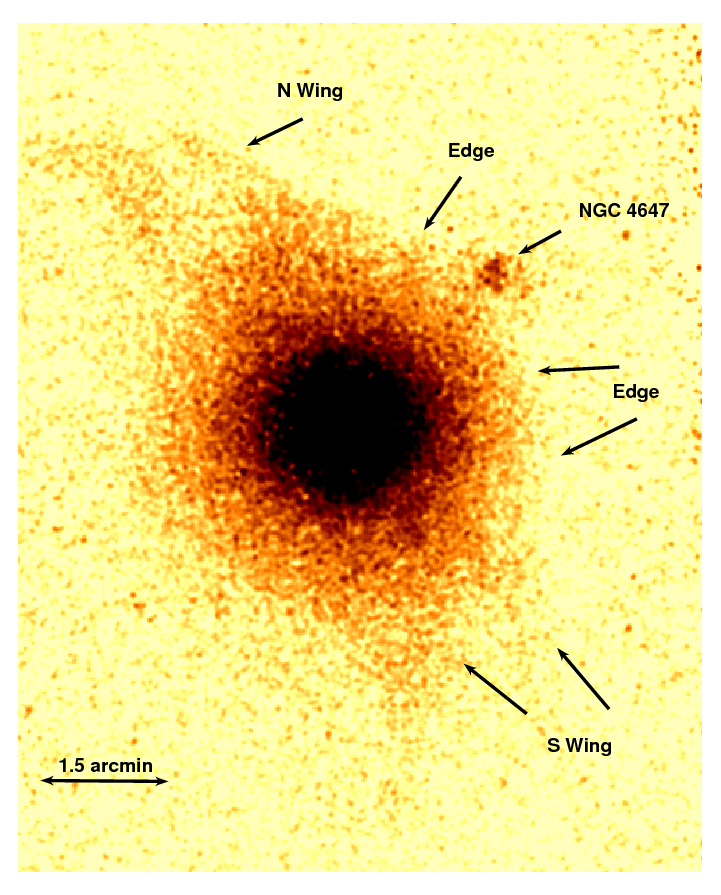}
\caption{Exposure-corrected, background-subtracted, 
  co-added  {\em Chandra} X-ray image of M60 in the soft band 
($0.5-2.0$\,keV). 
A bin size of $2 \times 2$ pixels ($1\as \times 1\as$) is used and the
image has been smoothed with a $2\as$ Gaussian kernel to highlight faint
diffuse features. Both the instrumental readout
and cosmic X-ray backgrounds have been subtracted and point sources have
been excluded. North is up and east is to the left. Note the
leading edge to the  north and west in the direction of M87 and gaseous wings
extending from the edge to either side. The small region of bright  
emission northwest of the surface brightness edge is the
spiral galaxy NGC\,4647.}
\label{fig:soft}
\end{center}
\end{figure}

M60 ($\alpha=12^{h}43^{m}39^{s}.6, \delta=11^{\circ}33^{'}09\as$), is
a massive elliptical galaxy located in projection 
$195$\,arcmin ($0.97$\,Mpc)  east of M87
($\alpha=12^{h}30^{m}49^{s}.4$, $\delta=12^{\circ}23^{'}28.0\as$). 
M60's line of sight velocity 
($v_{\rm{rad}}=1117\pm6$\kms; Trager \etal 2000)
compared to that of M87
($v_{\rm rad}=1307\pm7$\kms; Smith \etal 2000)
implies that M60 is moving through the Virgo ICM with at least 
$\Delta v_{\rm rad }=190\pm13$\kms towards us relative to M87. 
We obtain key measurements of the gas in the massive elliptical galaxy
M60 using archival {\em Chandra} X-ray observations (see Table
\ref{tab:M60ObsID}) with a cleaned 
coadded exposure time of $262$\,ks. These deep data sets reveal a wealth
of gas-stripping related structures, showing that M60 is undergoing
ram pressure stripping as it falls through the ICM towards the cluster
center. In the {\em Chandra} image (Fig.~\ref{fig:soft}), we see the
gas removed from the core of M60 has formed wings, a western edge and
a faint eastern filamentary tail. 
Understanding the nature of these edges and filamentary gaseous
structures and a determination of the velocity of M60 through the
Virgo ICM are the focus of this paper.

The discussion in this paper is presented as follows: In
\S\ref{sec:obs} we list the five {\em Chandra} observations used,
explaining the data reduction and processing
methods. \S\ref{sec:results} documents the critical steps undertaken
to analyze the {\em Chandra} data, highlighting the observational
results from the reprocessed, coadded data and discussing the results
drawn. Our conclusions are presented in \S\ref{sec:conc}. All
coordinates are J2000 and, unless otherwise indicated, all errors are
$90\%$ confidence levels. Using the Surface Brightness Fluctuation
(SBF) distances from Mei \etal (2007),
the distance to M87, is $17.2\pm0.6$ Mpc and $1^\prime=4.98$ kpc. 
These measurements assume a
Hubble constant $H_0=73$\kmsmpc in the $\Lambda$CDM
cosmology 
and are consistent with the SBF measurements by Tonry \etal (2001)
($16.1\pm1.2$\,Mpc).

%-----------------CHAPTER 2-OBSERVATIONS------------

\section{Observations and data reduction}
\label{sec:obs}

We reprocessed and then co-added five observations of M60
(Table~\ref{tab:M60ObsID}) taken with the {\em Chandra} X-ray
Observatory using the Advanced CCD Imaging Spectrometer array (ACIS)
with ACIS-S (chip S3) at the aimpoint, giving a total exposure of
$270$\,ks. A sixth, $38$\,ks  observation from the archive (ObsID 785) 
was badly flared and so was excluded from our analysis. All the data 
were taken in VFAINT mode and analyzed using the standard X-ray processing 
packages, CIAO 4.7 (CALDB 4.6.8), FTOOLS 6.15, Sherpa 4.4 and 
XSPEC 12.9.0. The data were filtered with 
{\em lc\_clean} and {\em deflare} to remove events in periods of 
abnormally low or high
counts, where the count rate deviates more than $20\%$ above or below
the mean. This gave a useful exposure time of $262$\,ks. 

\begin{deluxetable}{cccc}
\tablewidth{0pc}
\tablecaption{{\em Chandra} observations of M60}\label{tab:M60ObsID}
\tablehead{\colhead{ObsID} & \colhead{Date} & \colhead{Exposure} 
& \colhead{Cleaned Exposure} \\
& & (ks) & (ks) }
\startdata
8182 & 2007 Jan 30 & $52.37$ & $48.28$  \\
8507 & 2007 Feb 1 & $17.52$ & $17.52$ \\
12975 & 2011 Aug 8 & $84.93$ & $80.84$\\
12976 & 2011 Feb 24 & $101.04$ & $101.04$ \\
14328 & 2011 Aug 12 & $13.97$ & $13.97$ \\
\enddata
\tablecomments{The cleaned exposure is after removal of periods of 
anomalously high and low count rates.}
\end{deluxetable}

Some events from M60's bright core are redistributed 
along the ACIS readout direction during readout. These out-of-time events 
may contaminate both imaging and spectral analyses of the faint diffuse 
X-ray emission of interest. We model the readout contribution to the 
background using CIAO tool {\em readout\_bkg}, based on the algorithm 
developed by Vikhlinin \etal (2005).   The cosmic X-ray  
contribution to the background was taken from the blank-sky background (BSB)
files; a series of source-free background sets obtained at high
galactic latitude to avoid contamination from our Galaxy. For each
observation the relevant blank-sky background for that date was downloaded
and reprocessed. These were the period D source-free datasets for the
S3 CCD with exposure $400$\,ks. Identical energy filters and the same
cleaning process using CIAO were applied to the blank-sky background files. The
normalization of the background was set by the ratio of exposure times
between the M60 and blank-sky background datasets, with a last
adjustment to account for the time variability of the particle
background component made by matching  the detected count rate for 
background and source in the $10-12$\,keV energy
band, where particle backgrounds dominate. Both the readout and cosmic X-ray 
background contributions were subtracted from our subsequent image analysis.  
Exposure-corrected, background-subtracted flux images were created for 
each observation in the soft ($0.5-2.0$\,keV) and hard ($2.0-7.0$\,keV) 
X-ray energy bands, binned at two by two pixel resolution 
($0\farcs984 \times 0\farcs984$).
The  exposure-corrected, background-subtracted, coadded mosaic of the
five {\em Chandra} observations used in our analysis with the
identified point sources excluded is shown in 
Fig.~\ref{fig:soft}. The  dynamical gas features studied within this
paper are marked. The reader may view a less processed coadded mosaic of
the five data sets that includes  point sources in 
Fig.~\ref{fig:wings}.

%-------CHAPTER3-RESULTS-----------
\section{Results}
\label{sec:results}

M60 shows several features evident of gas
dynamics. Fig~\ref{fig:soft}, in the $0.5-2.0$\,keV energy band, shows
a surface brightness edge at $\sim 12$\,kpc north and west from the 
galaxy center in the direction of M87, a  north wing arcing to the
 northeast and a less prominent, possibly bifurcated, wing extending 
to the south.  
We will also show evidence for a faint diffuse tail downstream. 
These features are similar to those in M89 caused by
ram-pressure stripping as M89 interacts with the Virgo ICM during
infall (Machacek \etal 2006, Kraft \etal 2017). We suggest that M60 
is similarly undergoing gaseous stripping as it moves
through the Virgo ICM, forming the upstream edge, wings to the
northeast and south, and a faint diffuse eastern tail. 
A leading edge, faint tail and wings caused by Kelvin-Helmholtz
Instabilities (KHI) are expected for ram-pressure stripping through 
an inviscid medium (Roediger \etal 2015a).
We study the gas properties in these features in M60 and compare with 
the simulations of M89 (Roediger \etal 2015a,b), a similar Virgo galaxy,
to test this scenario. 

\begin{figure}[htb!]
\begin{center}
\includegraphics[width=3in]{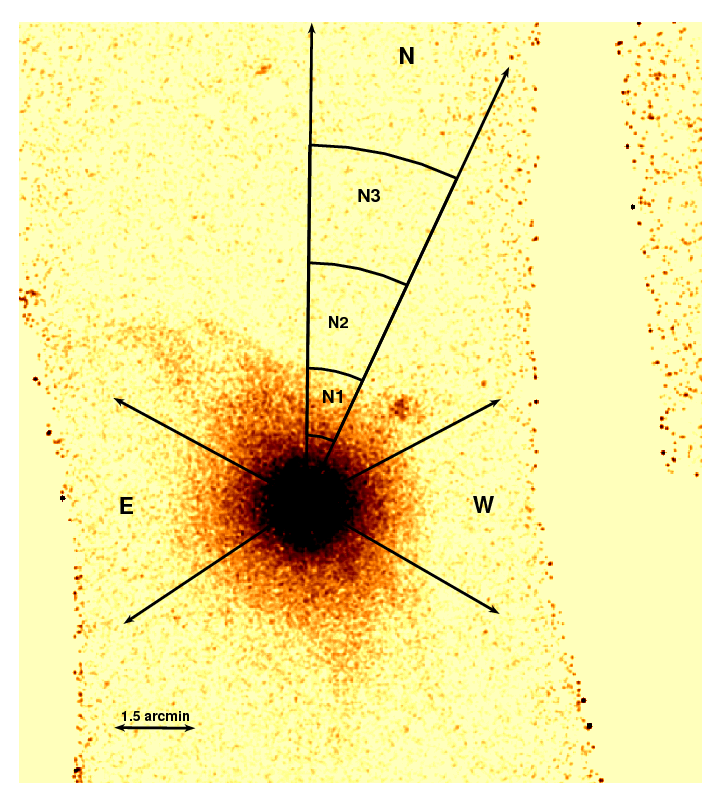}
\caption{Sectors for the profile analysis from 
  Table \protect\ref{tab:sectors} 
 overlaid on the background subtracted, exposure corrected $0.5-2.0$\,keV 
 Chandra image of M60 with point sources removed. 
 $1\,{\rm pixel}= 0\farcs984 \times 0\farcs984$.
Spectral regions along the N sector are also shown where N1 represents
galaxy gas inside the edge, N2 represents the ICM gas pile-up region,
and N3 is undisturbed Virgo ICM (free stream region). The (inner,
outer) radii of the spectral regions, measured from M60's center, are 
($75,150$), ($150,268$), and ($268, 400$) arcsec for N1, N2, and N3, 
respectively.}
\label{fig:edgereg}
\end{center}
\end{figure}

\subsection{Surface Brightness Profiles}
\label{sec:profiles}

To study the leading edge, we construct circular surface brightness 
profiles from the background subtracted, 
exposure corrected, coadded  $0.5-2$\,keV image, using logarithmic
radial steps.  The profiles are each centered on the X-ray peak of 
M60 ($\alpha=12^{h}43^{m}39^{s}.9, \delta=11^{\circ}33^{'}10.0\as$)
and confined to sectors to the north and west that avoid the wings and 
exclude the interacting spiral galaxy NGC\,4647. We similarly
construct the circular surface brightness profile in the eastern 
(downstream) direction to search for a diffuse, faint tail. These 
sectors are defined in Table \ref{tab:sectors}
and are shown in Fig. \ref{fig:edgereg}. 

\begin{deluxetable}{ccc}
\tablewidth{0pc}
\tablecaption{{Profile Sectors}\label{tab:sectors}}
\tablehead{\colhead{Profile Label} & \colhead{A} &
  \colhead{B} \\
  & (degree) & (degree) }
\startdata
 N  & $63$ & $88.3$ \\
 W  & $330$ & $27.6$ \\
 E  & $152$ & $214$  \\
\enddata
\tablecomments{The sector is centered at the X-ray peak and is defined as
  subtending the angle measured counterclockwise from angle A to angle
  B. All angles are measured counterclockwise from West. 
}
\end{deluxetable}

\begin{figure*}[htb!]
\begin{center}
\includegraphics[height=5 in, angle=270]{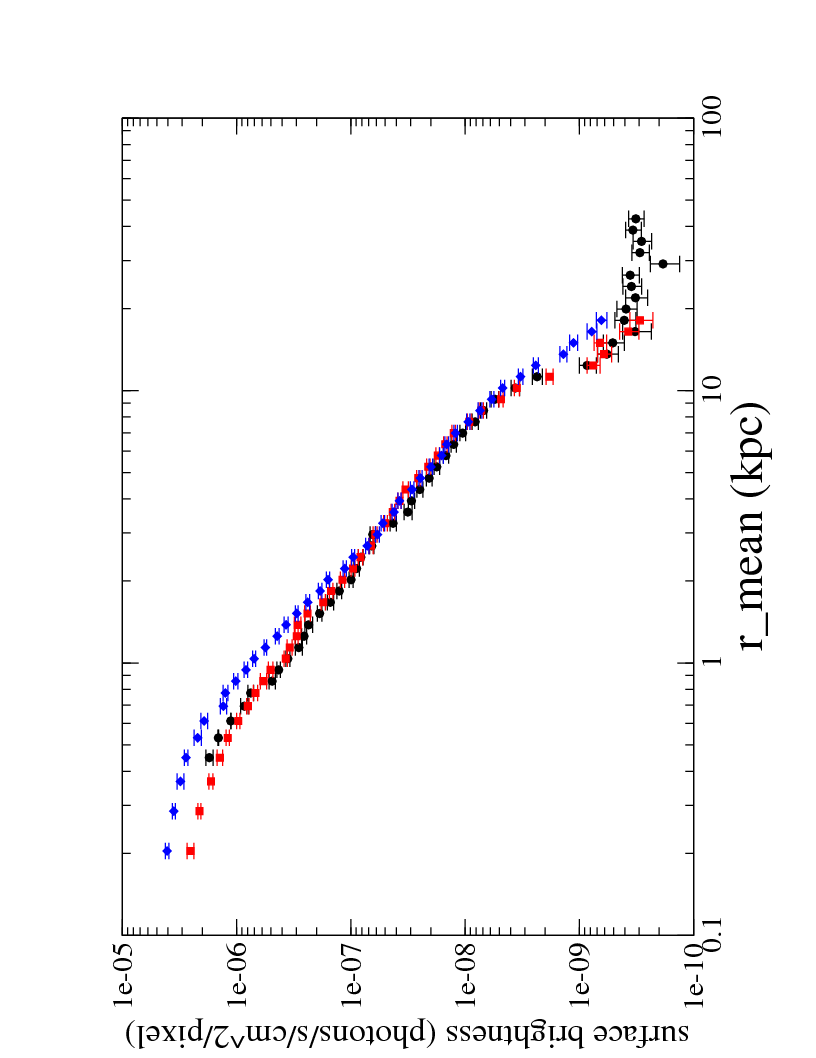}
\caption{Surface brightness profiles from the regions shown in 
Fig. \protect\ref{fig:edgereg}. The N  and W  profiles
(denoted by black circles and red squares, respectively) are the
same within uncertainties in their radial regions of overlap, beyond
which ($r > 22$\,kpc) the N profile flattens, consistent with that 
expected for Virgo emission. The E profile (blue diamonds) shows 
excess emission beyond $r \sim 11$\,kpc, consistent with a diffuse tail.}
\label{fig:profcompare}
\end{center}
\end{figure*}

Our results are shown in Fig. \ref{fig:profcompare}. Note that the
profiles in the N and W sectors are the same in the radial range of
overlap. This suggests that they are part of the same leading edge
caused by ram-pressure, only interrupted in the excluded region 
by the impending merger of NGC\,4647. Due to our Chandra coverage we
can study the morphology of the surface brightness profile to larger 
radii to the north (profile N) than to the west (profile W). 
We thus confine our analysis to the
profile of the N sector of the leading edge, shown in 
Fig \ref{fig:profcompare}, to 
measure gas properties to larger radii and probe farther from the
galaxy into the surrounding Virgo ICM. 
We see three distinct regions, the
characteristic edge profile for galaxy gas for 
$r < 149\as$\,($12$\,kpc), a very slowly varying, nearly flat 
surface brightness profile 
for $r> 268\as$\,($22$\,kpc), and a steeply falling region of excess
emission in between. At the $0.97$\,Mpc distance of M60 from M87, 
taken to be the center of the Virgo cluster, emission 
from undisturbed Virgo cluster gas would appear to be flat over 
the $\sim 50$\,kpc scale measured by the profile, consistent with the 
profile behavior at $r> 22$\,kpc. We take this to be representative of
the 'free stream region' from Vikhlinin \etal ( 2001). We suggest the
sharply falling surface brightness profile at $12 <r< 22$\,kpc may be
Virgo gas gravitationally attracted to M60's deep gravitational
potential at close radii, i.e. the pile-up region (Vikhlinin \etal 2001).  
These regions are also shown in Fig. \ref{fig:edgereg}. 

Vikhlinin \etal (2001)  showed the edge of a cool dense gas cloud 
moving through surrounding hot gas can be seen in radial X-ray surface
brightness profiles as a discontinuity (`edge') and that the relative
velocity of the gas cloud and its surrounding ICM can be determined by
the ratio of thermal gas pressures between gas at the stagnation point
and cluster gas in the free stream region. 
If the dense M60 gas were at rest relative to the ICM, the 
thermal pressures would be equal on both sides of the edge. However,  
for M60 infalling through the Virgo ICM, where it is
subjected to both ram and thermal pressures, this equality is only true at the
stagnation point where the relative gas velocity is zero. 
Following Vikhlinin \etal (2001), we use  
the gas pressure just inside the edge as a proxy for the pressure at
the stagnation point. We then use the pressure ratio
between gas at the stagnation point and in the free stream region to  
derive the relative velocity between M60 and the Virgo ICM, as
detailed in \S\ref{sec:M60infall}. 

To calculate pressures, characterize 
the nature of the edge, and, in the case of
ram pressure stripping, calculate the velocity of the galaxy, the 
relative temperatures and electron densities must be obtained for the 
gas on both sides of the edge.

\subsection{Gas Densities }
\label{sec:sbproffit}

As in Machacek \etal (2006), we assume spherically symmetric 
electron power law density models inside and outside the
edge  ($r=r_e$) 
\begin{equation}
  \begin{array}{l l}
	n_i(r<r_e)= & J_dn_0\left(\frac{r}{r_e}\right)^{-\alpha_i} \\
        n_o(r>r_e)= & n_0\left(\frac{r}{r_e}\right)^{-\alpha_o}
  \end{array}
\label{eq:densmod}
\end{equation}

with normalization $n_0$, inner and outer power law indices $\alpha_i$ and 
$\alpha_o$, respectively, and a 
discontinuous density 'jump' $J_d = n_i(r_e)/n_o(r_e)$ across the
edge.  We then determine the
density by integrating the X-ray emissivity along the
line of sight to fit the surface brightness profile, 
using a multi-variate $\chi^2$-minimization scheme with the
position of the edge ($r_e$), the density power law indices  
$\alpha_i$ and $\alpha_o$, and the square root of the 
surface brightness discontinuity $\sqrt{J_{sb}} \propto J_d$ across 
the edge allowed to vary. The measured surface brightness
discontinuity $J_{sb}$ is related to the density jump $J_d$ at the  
edge through the ratio of X-ray emissivities 
\begin{equation}
J_{sb} = \frac{\Lambda_i}{\Lambda_o}(J_d)^2
\label{eq:jump}
\end{equation}
where 
$\Lambda_i$($\Lambda_o$) are the cooling functions for gas
inside (outside) the edge, respectively. 

\begin{figure*}[htb!]
\begin{center}
\includegraphics[height=5 in,angle=270]{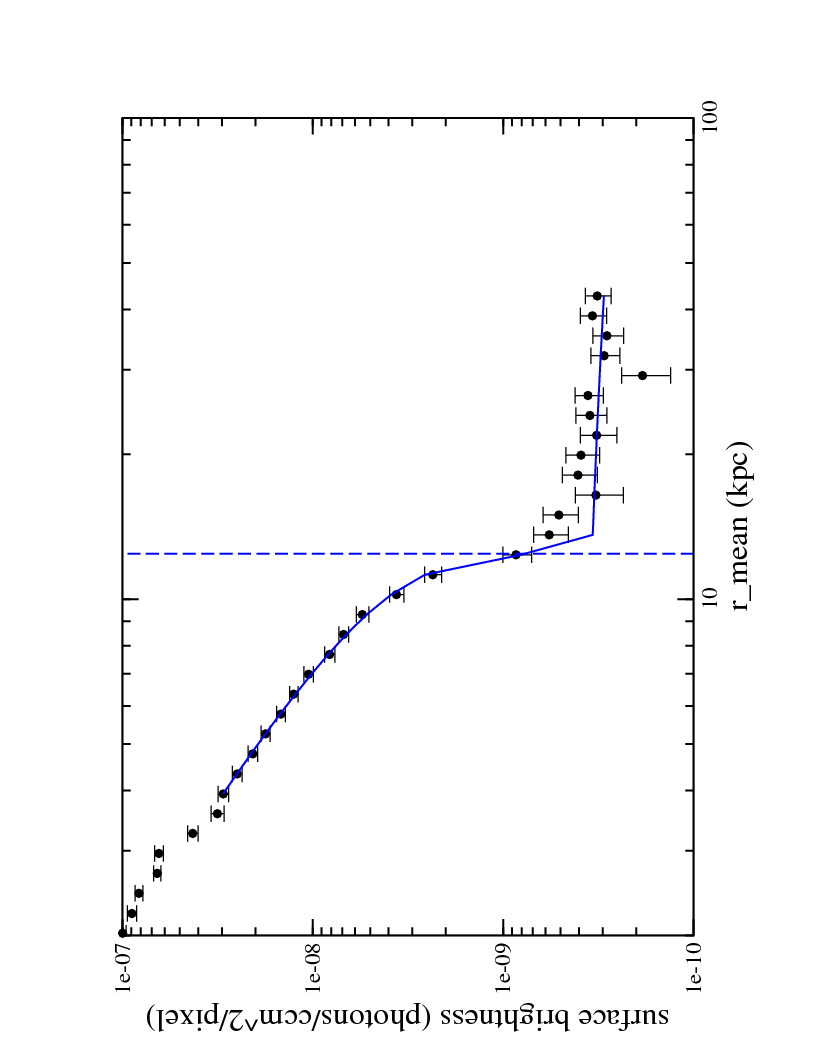}
\caption{ 
Fit to the radial surface brightness profile across the 
north edge. 
The vertical dashed line indicates 
the best-fit edge location. 
See Table \protect\ref{tab:edgefit}.}
\label{fig:Nedgeplot} 
\end{center}
\end{figure*}

At a projected distance of $\sim 1$\,Mpc from M87, we expect
the X-ray surface brightness of the Virgo ICM 
to be slowly varying 
over the angular scales probed by our merged observations. Thus we 
extrapolate the flat Virgo surface brightness at $r > 22$\,kpc to 
the observed edge and fit the resulting 
profile to determine
the gas density ratio between
galaxy gas inside the edge and undisturbed Virgo gas (the free stream
region) outside the edge. Our results
are shown in Fig. \ref{fig:Nedgeplot} and listed in Table
\ref{tab:edgefit}.
We find an edge position at $12.4\pm 0.1$\,kpc 
and $\sqrt{J_s} = 10.5^{+1.7}_{-1.1}$. 

For massive galaxy clusters where ICM gas temperatures are high, 
the cooling functions in Eq. \ref{eq:jump} are largely independent of gas 
abundances and only a weak function of the temperature such
that  $\Lambda_i/\Lambda_o \sim 1$  and  
$\sqrt(J_{sb}) \sim (J_d)$.  The density jump can be inferred  from
the surface brightness profile fits 
\clearpage
alone. However, 
for gas at lower temperatures ($\leq 1-2$\,kev), X-ray cooling is 
line dominated and the cooling function is strongly dependent on gas 
metallicity. If these abundances are different, which is likely the
case for galaxy gas and ICM gas in the outskirts of a galaxy cluster, 
the ratio of cooling functions in Eq. \ref{eq:jump} may deviate 
 significantly from $1$ and  must be 
determined from spectral modeling 
before we can infer the gas density ratio of interest and complete the
calculation of the infall velocity for M60. 
(see, e.g. Machacek \etal 2005).  

\begin{deluxetable}{cccc}
\tablewidth{0pc}
\tablecaption{{X-ray Surface Brightness Profile Model}\label{tab:edgefit}}
\tablehead{\colhead{$r_e$} &\colhead{$\alpha_i$} &
  \colhead{$\alpha_o$} & \colhead{$\sqrt{J_{sb}}$} \\
   (kpc) &  &  &  }
\startdata
 $12.4\pm 0.1$ & $-1.29^{+0.07}_{-0.08}$ & $-0.38^{+0.09}_{-0.05}$   
  & $10.5^{+1.7}_{-1.1}$ \\
\enddata
\tablecomments{ The density model extrapolates the slowly varying
  Virgo ICM over the pile-up region to the edge. 
Columns are edge location $r_e$, power law indices
$\alpha_i$($\alpha_o$) inside and outside th edge, and square root of
the surface brightness jump. The density model extrapolates the slowly varying
  Virgo ICM over the pile-up region to the edge. }
\end{deluxetable}

\subsection{Spectral Modeling: Gas Temperatures and Abundances}
\label{sec:spectra}

To determine the mean spectral properties of the gas in M60, we use CIAO
tool {\em specextract} to extract a mean spectrum in a $200\as$
circular region centered on the observed X-ray peak. We also extract
spectra for the regions N1, N2, and N3 along the northern profile shown in 
Fig. \ref{fig:edgereg} to characterize the Virgo emission (N3) at the
position of M60 and to complete the cold front analysis across the
leading surface brightness edge (see Fig. \ref{fig:Nedgeplot}). Point
sources above a detection threshold of $\sim 2 \times 10^{37}$\ergs
(see Appendix), as well as emission from the interacting companion 
galaxy NGC\,4647, are excluded from the data. 
The resulting spectra are modeled using XSpec 12.9.0 (See Arnaud 1996).   

\subsubsection{Mean X-ray Spectral Properties of M60}
\label{sec:M60mean}

We first consider the average spectral properties of M60 as a whole
using the blank sky background files as in \S\ref{sec:profiles} for 
backgrounds. This will allow us to determine the mean abundance as
well as temperature of gas within M60. ACIS read-out effects 
redistribute $1.3\%$ of the source 
counts along the read-out direction. Since most of these photons are
from the bright core, which is included in our mean spectrum, and 
inspection of the read-out map shows that only $0.3\%$ of the total 
source counts are redistributed outside the $200\as$ circular spectral 
region, read-out will not significantly affect the mean spectral fit. 
We use an absorbed VAPEC model, that assumes X-ray emission from
collisionally ionized diffuse gas  with emission rates calculated from
the most current atomic transition data tabulated in 
AtomDB.\footnote{http://atomdb.org}  We fix the hydrogen column density at
the Galactic value ($N_{\rm{H}}=2.1 \times 10^{20}$\cms). For more
information on the VAPEC model, please see {\it XSpec: An X-ray Spectral
Fitting Package User's Guide} by Arnaud, Gordon, and 
Dorman.\footnote{https//heasarc.nasa.gov/docs/xanadu/spec/manual/manual.html}

The temperature and abundances for Fe, Mg, Si, and O were allowed 
to vary. All other abundances were fixed at solar. We find a mean
temperature $kT=0.906 \pm 0.004$\,keV with  Fe, O, Mg and Si Anders
\& Grevesse (1989) abundances
of $0.46\pm 0.2\Zs$, $0.16\pm 0.04\Zs$, $0.93^{+0.05}_{-0.06}\Zs$, and
$0.91 \pm 0.05\Zs$ for $\chi^2/({\rm dof}) = 1989/1516$.  Total intrinsic
fluxes and luminosities for M60 for the assumed $17.2$\,Mpc distance
are listed in Table \ref{tab:200lum}.   

Although a thermal plasma VAPEC model provides an excellent fit to
the data, {\em Chandra} data measure the X-ray emission from all sources in
M60 and along the line of sight. This includes X-ray emission from 
any unresolved point sources, such as
cataclysmic variables (CV's), accreting white dwarfs (AB's) and low
mass X-ray binaries (LMXBs), and Virgo ICM emission along
the line of sight, as well as the diffuse galaxy gas. 
 We model the relative contribution of
each of these  unresolved stellar X-ray sources in the Appendix to
determine whether these sources significantly affect our spectral measurements
of flux, temperature and metallicity of M60's diffuse gas.  
We find that CV's and AB's contribute only  $1.5\%$   and LMXB's
$2\%$ of M60's total $0.5-2$\,keV flux. Thus M60 is still gas
dominated. 
If these stellar components are included in our spectral model, our 
results for the temperature and 
abundance of the diffuse galaxy gas are 
unchanged. Furthermore if we allow the normalization of the power law
model for unresolved LMXBs to be a free parameter, the measured 
$0.5-2$\,keV flux of unresolved 
LMXBs in this region is consistent with our predicted value.

\begin{deluxetable}{cccc}
\tablewidth{0pc}
\tablecaption{{M60 Intrinsic X-ray fluxes and luminosities}\label{tab:200lum}}
\tablehead{\colhead{Band} & \colhead{Range} & \colhead{Flux} 
& \colhead{Luminosity} \\
 & (keV) & $(10^{-12}$\ergscm) & ($10^{40}$\ergs)}
\startdata
Soft & $0.5 - 2.0$ & $3.27$ & $11.56$ \\
Hard & $2.0 - 7.0$ & $0.27$ & $0.96$ \\
Full & $0.5 - 7.0$ & $3.54$ & $12.53$ \\
\enddata
\tablecomments{ Flux and luminosities from the best absorbed VApec
  model fit to the mean spectrum of M60 using a $200\as$ radius 
  circular region centered on M60's X-ray peak.}
\end{deluxetable}

\subsubsection{Virgo ICM}
\label{sec:virgospec}

The final component to model in fitting the spectra is the
contribution from the cluster emission. Virgo is a young, 
dynamically active cluster.  Urban \etal (2011) found from {\it XMM-Newton}
observations that beyond $500$\,kpc from M87, the X-ray surface
brightness of the Virgo ICM is highly variable, varying as much as a
factor $2$ at $0.9$\,Mpc, the projected distance of
M60 from M87. Using Suzaku, Simonescu \etal (private communication) 
further showed that the flux
and temperature of Virgo gas at large radius depends strongly on the azimuthal
direction of the observation, with higher fluxes and temperatures
found along directions that coincide with higher density filaments. 
We thus choose to measure the spectral properties of the Virgo gas 
using our combined Chandra data in region N3 (see
Fig. \ref{fig:edgereg}), 
where the northern
surface brightness profile shown in Fig. \ref{fig:profcompare} flattens, rather
than adopt a model from the literature. 
We model the spectrum of the Virgo ICM in region N3 using an absorbed 
thermal plasma APEC model with fixed Galactic hydrogen absorption 
and $0.1\Zs$ abundance (Urban \etal 2011). We find best-fit 
temperature, model normalization and intrinsic $0.5-2$\,keV flux of 
$1.37^{+0.35}_{-0.19}$\,keV, $1.75 \times 10^{-5}$\,cm$^{-5}$, and 
$8.9 \times 10^{-15}$\ergscm (see Table \ref{tab:nedgespec}). 
X-ray emission in the cluster outskirts is faint, contributing only
$0.3\%$ of the $0.5-2$\,keV flux in the $200\as$ circular region
emcompassing M60. From the VAPEC model normalization we find a
  mean electron density for this region of $3 \times 10^{-5}$\cmc with
  $\sim 30\%$ uncertainties, yielding a thermal gas pressure of 
$\sim 1.35 \times 10^{-13}$\,ergs\cmc. These results are consistent
with Urban \etal (2011), given the measured variability  observed at 
these large radii. 
Since we directly measure the thermal properties of the Virgo
  cluster gas, i.e. gas density, temperature and thermal gas pressure, 
from the spectral fitting of the X-ray emission of the gas, any
evolutionary effects of the ambient cluster magnetic field on
determining those thermal properties have implicitly been accounted
for in that measurement. Ambient magnetic fields could also  
contribute an additional non-thermal component to the total pressure
in the Virgo Cluster. However,  
using the simulated magnetic field profile for the Virgo Cluster from Pfrommer
\& Dursi (2010), the expected non-thermal pressure due to ambient 
magnetic fields is at most $\sim 1\%$ of the thermal pressure at 
these radii and likely could be much less. Thus it does not
significantly affect our analysis. 
 
\begin{deluxetable*}{cccccc}
\tablewidth{0pc}
\tablecaption{{Northern Edge Spectral Models}\label{tab:nedgespec}}
\tablehead{\colhead{Region} & \colhead{$kT$} & \colhead{norm}
  &\colhead{flux} &\colhead{$\Lambda$} & $\chi^2/({\rm dof})$\\
 & (keV) & ($10^{-5}$\,cm$^{-5}$) & ($10^{-14}$\ergscm) &
 $10^{-23}$\,ergs\,cm$^3$\,s$^{-1}$) & }
\startdata
N1 &$1.00 \pm 0.02$ &$2.34$ &$3.19$ &$1.36$ &$112/127$ \\
N2 &$1.6^{+0.5}_{-0.3}$ &$1.82$ &$0.93$ &$0.51$ & $186/154$\\
N3 &$1.37^{+0.35}_{-0.19}$ &$1.75$ &$0.89$ &$0.51$ & $293/243$\\
\enddata
\tablecomments{Spectra were modeled using an absorbed Apec model with
Galactic absorption and abundance fixed at $0.5\Zs$ for M60 galaxy 
gas (N1) and at $0.1\Zs$ for regions N2 and N3 consistent with Virgo 
Cluster gas}
\end{deluxetable*}

\subsubsection{Spectral Fitting Across the Edge}
\label{sec:edgespecfit}

To determine the infall velocity of M60 into the Virgo galaxy cluster, 
we need to measure the temperatures and densities of gas on either
side of the edge. We chose the northern profile sector for our analysis
because detector coverage allows us to measure gas properties out to
greater distances consistent with undisturbed Virgo gas (N3).
Additionally the northern sector lies  perpendicular to the readout 
direction, and the inner radius of region N1, the region just inside the 
edge, lies outside the bright core region, where readout would be 
greatest, minimizing the contribution of these out-of-time events in
these regions.  Thus readout in the spectral analysis of the northern regions
(N1,N2,N3) may be neglected. We also note that a  comparsion of X-ray
and kinematical mass measurements for M60, show that magnetic fields do
not contribute significantly to the pressure inside the edge in region
N1 ($6.2 <r <12.4$\,kpc (Paggi \etal 2014).

We model the inner two northern regions (N1 and N2), similar to that of N3 in
Section \ref{sec:virgospec},  with an absorbed APEC model
with fixed Galactic hydrogen absorption and abundances at $0.5\Zs$ for
region N1 in  M60, consistent with our measured  mean Fe abundance for M60
galaxy gas, and at the Virgo value ($0.1\Zs$) for gas in the
pile-up region N2. Our results are given in Table
\ref{tab:nedgespec}. 

Since both galaxy and Virgo gas have temperatures $\sim 1$\,keV,  
the cooling function $\Lambda(A,T)$ and thus emissivity 
$\Lambda n_en_p$ for each depends sensitively on the metal 
abundance $A$ of the gas. 
We determine the cooling functions for each of the northern spectral
regions (N1, N2, N3) using  
\begin{equation}
 \Lambda = \frac{10^{-14}FD_L^2}{N_{\rm{APEC}}[D_A(1+z)]^2} 
\label{eq:lambda}
\end{equation}
where $F$ is the unabsorbed model flux in the $0.5-2$\,keV energy band, to
match that of the surface brightness profile, $N_{\rm{APEC}}$ is the
APEC model normalization, $z$ is the redshift, $D_L(D_A)$ are the
luminosity (angular size) distances, respectively, and $D_L^2 \sim
[D_A(1+z)]^2$ for $z << 1$. Values for 
$N_{\rm{APEC}}$, $F$, and $\Lambda$ 
are also given in  Table \ref{tab:nedgespec}. 

\subsection{Constraining M60's dynamical motion} 
\label{sec:M60infall}

Using the fitted surface brightness jumps $\sqrt(J_{sb})$ from Table
\ref{tab:edgefit} and the cooling functions given in Table
\ref{tab:nedgespec} in Equation \ref{eq:jump}, we calculate the density
ratio between  galaxy gas inside the edge and free streaming Virgo gas 
and multiply  these by the respective temperature ratios from the
spectral fits to obtain the pressure ratio between the stagnation
point and the Virgo free stream region. Uncertainties in derived
ratios are estimated using the extreme values of the $90\%$ CL 
uncertainties for measured properties. Our results
are listed in Table \ref{tab:ratios}.  

\begin{deluxetable*}{cccccccc}
\tablewidth{0pc}
\tablecaption{{M60 Velocity Analysis}\label{tab:ratios}}
\tablehead{ \colhead{$T_i/T_o$} & \colhead{$\Lambda_i/\Lambda_o$} &
  \colhead{$n_i/n_o$} &\colhead{$p_i/p_o$} &\colhead{ $M_a$}
  &\colhead{ $v$} &\colhead{$v_t$} &\colhead{$\xi$} \\
 &  &  & & & (\kms) & (\kms) & (deg) }
\startdata
$0.73^{+0.13}_{-0.17}$ &$2.66$ &$6.44^{+1.04}_{-0.67}$ & $4.7^{+1.7}_{-1.4}$ 
&$1.7 \pm 0.3$ &$1030 \pm 180$  &$1012^{+183}_{-192}$  &$11\pm 3 $ 
\enddata
\tablecomments{Velocities assume a sound speed of $604$\kms 
for  $1.37$\,keV 
Virgo gas and M60 radial velocity $v_r=-190 \pm 15$\kms. 
Uncertainites for derived values assume extremes in the 90\% CL 
uncertainties for measured properties.}  
\end{deluxetable*}

The Mach number $M_a=v/c_s$ (where $c_s$ is the speed of sound in the
cluster free stream region) for the cold gas cloud moving through the 
hot ICM is determined from the ratio of pressures (see
Eqs. \ref{eq:pratiosub} and Eq. \ref{eq:pratiosuper} ).

\begin{equation}
	\frac{p_i}{p_o}=\left(1+\frac{\gamma - 1}{2} M_a^2 \right)^{\frac{\gamma}{\gamma -1}}, M_a \le 1
\label{eq:pratiosub}
\end{equation}

\begin{equation}
	\frac{p_i}{p_o}=\left(\frac{\gamma+1}{2}\right)^{\frac{\gamma+1}{\gamma-1}} M_a^2 \left(\gamma - \frac{\gamma-1}{2M_a^2}\right)^{\frac{-1}{\gamma-1}}, M_a \ge 1
\label{eq:pratiosuper}
\end{equation}

where $\gamma=5/3$ is the adiabatic index of the monatomic gas.

From the pressure ratios given in Table \ref{tab:ratios} and 
Eqns. \ref{eq:pratiosub} and \ref{eq:pratiosuper}, we find M60 moving
at Mach $1.7 \pm 0.3$ relative to the Virgo ICM. 
The sound speed in completely ionized  $1.37$\,keV 
Virgo gas in the free stream region is $c_s= 604$\kms, such that  
we find the speed  $v=M_ac_s$ of M60 
relative to the Virgo ICM is $1030 \pm 180$\kms.

The physical separation of M60 and M87 is 971 kpc, confirmed with the 
distance moduli measurements in both Tonry \etal (2001) and  
Mei \etal (2007). The radial velocity difference 
between M60 and M87 is $\Delta v_{\rm{rad}}=-190\pm15$\kms 
(Trager \etal 2000). Taking the Virgo cluster ICM to be at 
rest relative to M87, we determine the relative transverse component 
of M60's velocity as $v_t=1012^{+183}_{-192}$\kms. 
The components of M60's motion through the Virgo ICM imply an inclination 
angle with respect to the plane of the sky of $\xi=11 \pm 3$ degrees.
Nearly all of M60's motion is therefore in the plane of the
sky, consistent with the sharp surface brightness edge observed. We
estimate the time $t_{pp}$ for pericenter passage
assuming a constant infall velocity and direct projected path towards
M87, such that $t_{pp} \sim 0.971 Mpc/v_t= 0.95$\,Gyr.  If the 
motion of M60 in the plane of the sky is directly towards M87, this
will be an upper bound on the infall time. 

  MHD simulations suggest that magnetic fields may wrap around a
  galaxy during infall, forming a thin draping layer. In this layer
  ambient cluster magnetic fields may become amplified to a maximum
  value such that their magnetic pressure equals the ram pressure of
  the ICM on the galaxy (Lyutikov 2006; Pfrommer \& Dursi
  2010). Recent simulations also suggest these fields may become
  tangled and not suppress ram pressure stripping (Ruszkowski \etal
  2014). For M60 infalling at $1030$\kms through Virgo gas of electron
  density $3 \times 10^{-5}$\cmc, this would imply a possible maximum
  magnetic field of $\sim 3.9\,\mu{\rm G}$ within a $1\as.5$
  ($120$\,pc) thick draping layer outside the edge. Unfortunately our
  combined observations of M60 are not deep enough to constrain the
  properties of any such layer (See, e.g. Su \etal 2017 for the
  analysis of NGC\,1404, a similar galaxy, but with a {\it Chandra}
  exposure a factor $2.6$ longer than the one presented here).

\begin{figure}[htb!]
\begin{center}
\includegraphics[width=3in]{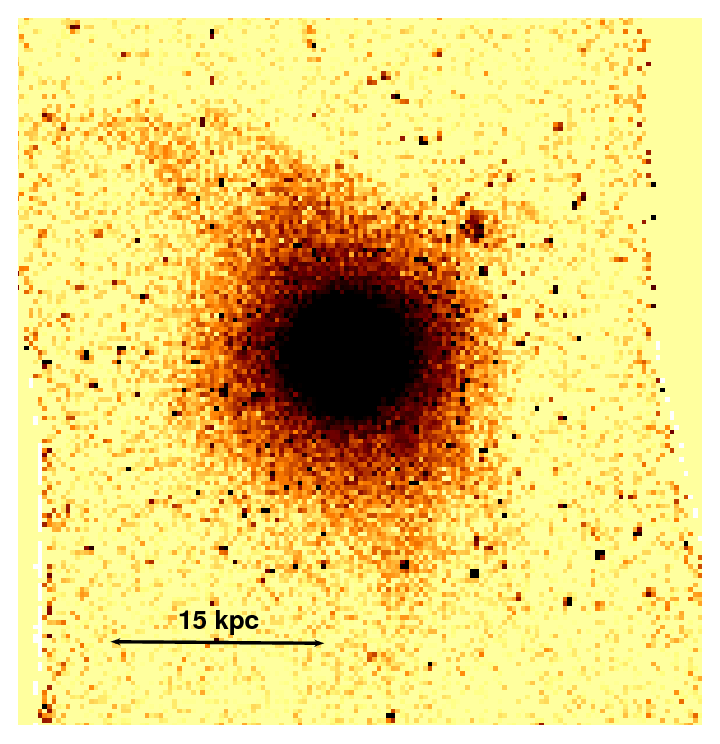}
\includegraphics[width=3in]{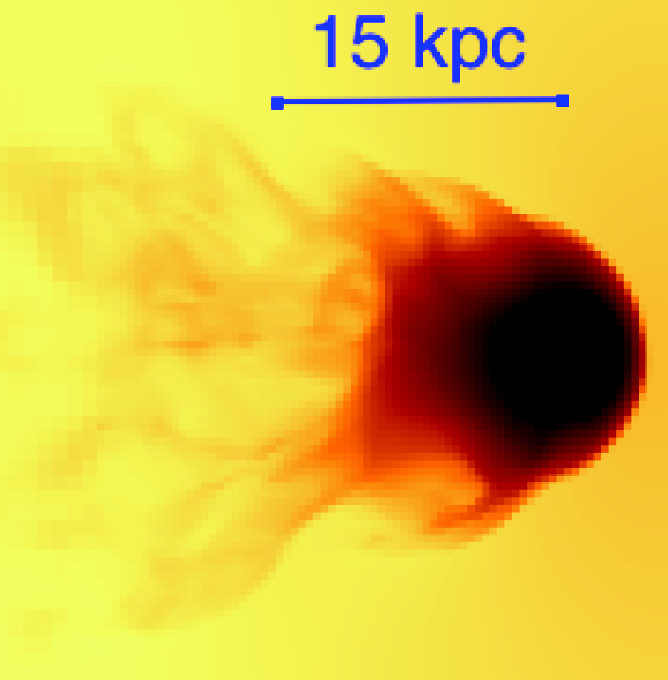}
\caption{(upper) Coadded {\em Chandra} image of M60 with the double
          wing structures highlighted by blocking to a bin size of 
   $8\times8$ pixels ($\sim4^{\as}\times4^{\as}$).
(lower) Simulation of gas stripping and Kelvin-Helmholtz 
    instabilities in M89, which form "wings" similar to those in the 
   {\em Chandra} image of M60. The frame is a low energy band 
   ($0.5-2.0$\,keV) X-ray projection.} 
\label{fig:wings}
\end{center}
\end{figure}

\subsection{Wings and Tail Structures of M60} 
\label{sec:wingstail}

The motion of M60 through the Virgo ICM 
causes the gas on the
northwest side to be stripped and pushed behind the galaxy in the wake of 
its passage through Virgo, forming the observed `wings' and sharp northwest 
edge. The coadded images show there are a pair of wings in both the
northeast and south directions. These wings align with the direction of
the outburst of the AGN at the core of M60 (Paggi \etal 2014). 

The double wing structure is shown more clearly by binning to 
$8\times8$ pixels ($4^{\prime\prime}\times4^{\prime\prime}$ bins) 
in Fig~\ref{fig:wings}. 
The filamentary wings are thin and long, extending at least 
$150^{\prime\prime}$ in 
both directions and as wide as the galaxy atmosphere radius 
($150^{\prime\prime}$  span across each double wing structure). 
The northeastern wings are of approximately equal length
($150^{\prime\prime}$) 
and separated by  a cavity $28^{\prime\prime}$ wide. The wings to the 
south are less symmetric, with the front wing being shorter than the 
downstream wing by a factor of $2$.

The dimensions of these filamentary wings were compared with
simulations by Roediger \etal (2015a,b) of the Virgo galaxy M89 (NGC4552). 
The lower panel of Fig~\ref{fig:wings} shows a soft band
X-ray projection of M89 that demonstrates how turbulent stripping can
create wings of diffuse gas to the sides of the galaxy, as we observe
for M60. The size of the wings scales with the size of the gas
atmosphere, hence larger wings are seen in M60, compared to the
simulations of the Virgo elliptical M89, since most of M89's gas has
already been stripped.

The double wing structure is seen in the simulation image 
(Fig~\ref{fig:wings} lower panel). The wings are attached directly to the 
sides of the galaxy as is the case for M60, another indicator of motion 
nearly in the plane of the sky. This can also be explained by the 
inclination angle; the simulations assume motion purely in the plane 
of the sky, while M60 has an inclination of $11 \pm 3$ degrees as
shown in \S\ref{sec:M60infall}. For M89 these features are attached to 
the upstream edge rather than to the sides (Machacek \etal 2006). 
Thus the simulation better represents M60 in this instance.  

Two processes can disturb the sharp-edge of a cold front; 
Rayleigh-Taylor instabilities  and Kelvin-Helmholtz (KHI) instabilities 
(Roediger \etal 2012). The Rayleigh-Taylor instability occurs 
when two fluids of different densities are accelerated into each
other. The galaxy gas is accelerating into the ICM, but the 
Rayleigh-Taylor instability  is suppressed by M60's gravity, 
so that the effective acceleration points in the opposite direction.

KHI instabilities, formed at the surface of the cold front edge where
velocity shears may produce  turbulent motions, are likely the cause 
of the filamentary wings seen in Figs. \ref{fig:soft} and
\ref{fig:wings}, perhaps aided by previous AGN outbursts that may 
have  lifted  the galaxy gas upward in M60's gravitational potential 
making it easier to strip.

The simulations predict some faint emission behind the wings
downstream leading to a diffuse tail trailing the galaxy. There is a
faint excess of emission seen for $r> 10$\,kpc in the eastern surface
brightness profile (Fig. \ref{fig:profcompare}) over that observed in
the northern or western profiles, suggestive of such a tail.
 
The simulations of M89 predict a downstream edge marking the boundary
between the galaxy atmosphere and diffuse tail. 
We performed our edge
analysis as outlined in \S\ref{sec:sbproffit}, on the eastern profile 
of M60 to look for this downstream edge (See Figs. \ref{fig:edgereg}
and \ref{fig:profcompare} and Table \ref{tab:sectors}).

\begin{figure}[htb!]
\begin{center}
\includegraphics[height=3 in, angle=270]{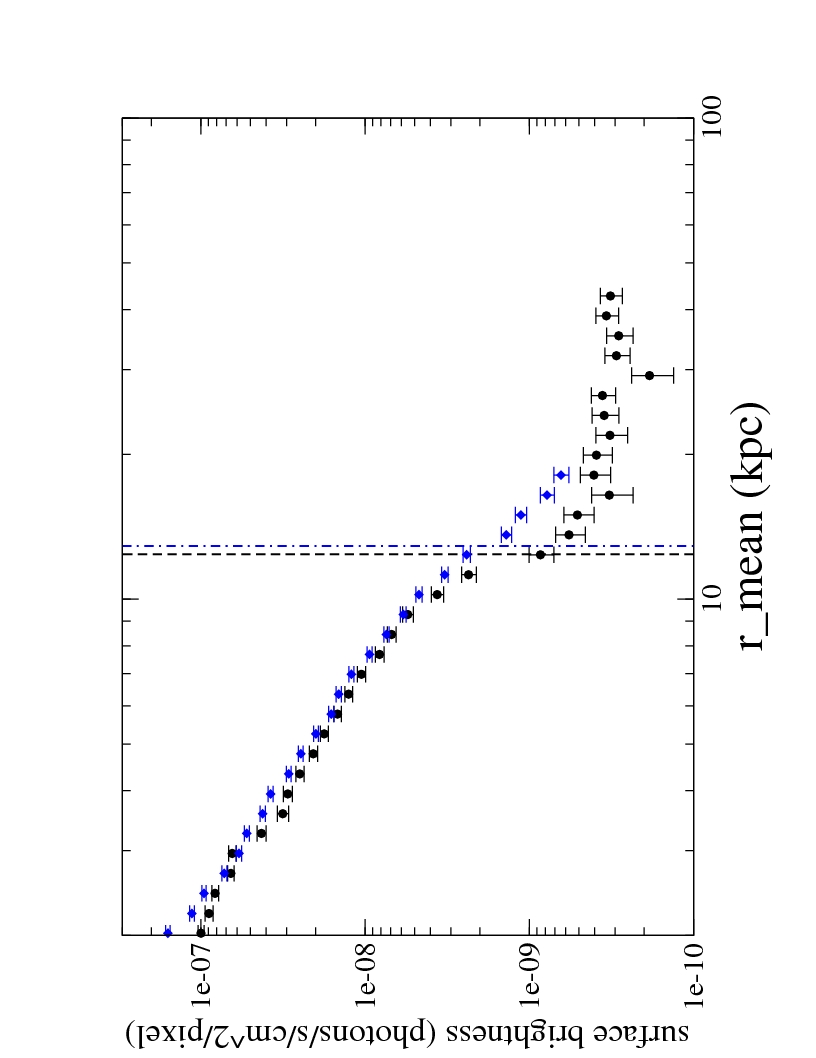}
\caption{Surface brightness radial profile for the northern upstream 
(black dots)
and eastern downstream (blue diamonds) regions of M60. 
Note the excess emission to the east 
at $r>10$\,kpc  suggesting a faint, diffuse gas tail. Vertical lines
denote the edge locations for the upstream northern profile (black
dashes) and the eastern downstream tail (blue dot-dashed).} 
\label{fig:tailcompare} 
\end{center}
\end{figure}

Using the power law model from Eq. \ref{eq:densmod}, the eastern
profile is well fit with upstream and downstream power law slopes of 
$\alpha_i=1.42^{+0.03}_{-0.04}$
and $\alpha_o=1.86^{+0.38}_{-0.32}$, respectively. We find a small
jump $\sqrt(J_{sb})=1.43^{+0.23}_{-0.17}$ at
$r=12.9^{+0.4}_{-0.5}$\,kpc downstream from the center of M60. 
Since we expect from simulations that the near  tail is composed of 
galaxy gas displaced but not yet completely stripped from M60
(Roediger etal 2015a), the abundance across the eastern edges should 
be the same. Thus $\sqrt(J_{sb}) = J_d$, the ratio
of the gas density across the eastern downstream edge. 
Fig~\ref{fig:tailcompare} shows the extracted surface
brightness profiles for both the upstream (northern cold front) and downstream
(eastern tail) edges. The simulations generally find the downstream edge at a
larger radius than the upstream edge, even more so than is observed
for M60. 

The surface brightness profiles are symmetric in both radial
directions until each respective edge. In the simulations of M89, 
the upstream edge of the cold front and wings look sharper than 
the downstream edge, consistent with what we see in 
Fig~\ref{fig:tailcompare}.  

There are several factors that could keep the tail of M60 faint. A
steep initial gas density profile for M60 would mean there was less
gas to strip at larger radii and the stripped gas may mix quickly with
the surrounding ICM. The large distance of M60 from M87 ($971$\,kpc)
and the galaxy velocity of $v_{\rm{M60}}= 1030 \pm 180$\kms
suggest multiple possibilities. First, we could be witnessing
the very early stages of the M60-Virgo interaction, thus sufficient
gas has not yet been stripped to form a clear tail as observed in M89,
which is located far closer to M87 ($390$\,kpc) and therefore has undergone a
much longer gas stripping period.  Alternatively, M60 may already have
passed through the Virgo system once and we are observing it shortly
upon completing a turn around in its orbit. This would suggest
stripping is occurring primarily from KHI, such that no large volumes of gas
are being pushed behind the galaxy. In either case, stripping may be less
efficient in the low density Virgo outskirts, requiring gas to first
be uplifted by periodic AGN activity before being pushed back into the
tail. 

%---------------------------------------CHAPTER4-CONCLUSION--------------------------------------------

\section{Conclusions}
\label{sec:conc}

Using archival data from the {\em Chandra} X-ray Observatory (total
cleaned exposure time $262$\,ks), we identified a surface brightness 
discontinuity (edge) $r_e=12$\,kpc from the center of M60 
to the north and west in the direction on M87, the Virgo Cluster
center. The surface brightness edge is produced by the ram pressure stripping
of the gas in M60 as it passes through the Virgo ICM. From the
surface brightness profile, taken to the north to minimize instrumental
effects and maximize radial distance coverage, we measured gas temperatures 
within the edge in M60, in the Virgo gas pile-up
region  ($12.4\,{\rm kpc} < r < 22\,{\rm kpc}$) immediately outside the edge,
and in the Virgo free stream region ($r>22$\,kpc), and measured the 
density ratio between gas  inside the edge and
in the Virgo free stream region to determine the physical motion of
the elliptical galaxy M60 through the Virgo ICM. 

We find:  
\begin{itemize}

\item{X-ray emission in M60 is gas dominated and well fit in the mean
    by a VAPEC spectral model with $kT = 0.906 \pm 0.004$\,keV and Fe,
    O, Mg, and Si abundance of $0.46 \pm 0.2\Zs$, $0.16 \pm 0.04\Zs$,
$0.93^{+0.05}_{-0.06}\Zs$, and $0.91\pm0.05\Zs$, respectively.}

\item{ Fixing the abundance inside the edge at $0.5\Zs$ consistent
with M60's mean results and at $0.1\Zs$ for Virgo gas at $0.971$\,Mpc
from the literature, gas temperatures along the northern profile for galaxy gas
inside the edge, in the pile-up region, and in the Virgo free stream
region are $1.00\pm 0.02$\,keV, $1.6^{+0.5}_{-0.3}$\,keV, and
$1.37^{+0.35}_{-0.19}$\,keV, respectively.} 

\item{The leading (northern) 
edge at $r_e=12.4\pm 0.1$\,kpc from the galaxy center,  coincides with a
jump in density of $n_i/n_o=6.44^{+1.04}_{-0.67}$
between gas inside the edge and Virgo gas in the free stream region.}

\item{ The measured pressure ratio $p_i/p_o=4.7^{+1.7}_{-1.4}$ between
galaxy gas inside the edge and in the Virgo free stream region implies
M60 is infalling with total velocity $v_{\rm{M60}}=1030 \pm 180$\kms 
(Mach $1.7\pm 0.3$) relative to the Virgo ICM. 
Given the relative radial velocity of 
$\Delta v_{\rm{rad}}=-190 \pm 15$, this yields an inclination angle
$\xi=11 \pm 3$\,degrees with respect to the 
plane of the sky, consistent with the observed sharp surface
brightness edge. M60's inferred transverse infall velocity, 
$v_t=1012^{+183}_{-192}$\kms, places an upper bound on the time to 
pericenter passage of $\sim 0.95$\, Gyr.}

\item{Extended wing-like features are observed to the northeast and
    south of M60. Comparison of these features with simulations of
    ram-pressure stripping suggests that these wings are likely
    produced by Kelvin-Helmholtz instabilities caused by M60's 
    motion through a nearly inviscid Virgo ICM. 
The thin and long filamentary gas wings scale with the size of the gas 
atmosphere and are attached directly to the sides of M60, an indicator
of motion in the plane of the sky, consistent with the small measured
inclination angle.} 

\item{Excess emission observed to the east (downstream) of M60's
    motion and confirmed in the eastern surface brightness profile is
    consistent with the existence of a faint, diffuse tail, similar to
    those seen in simulations of ram pressure stripping. The 
faintness of the tail is due either to insufficient stripping of the 
gas due to M60's distance from the cluster core, or because we are observing 
M60 shortly after completing a turn-around of its orbit.}

\end{itemize}

\section*{Acknowledgments}\label{sec:ack}

Data reduction and analysis was supported by the CXC CIAO, Sherpa, 
XMM-ESAS software packages and CALDB v4.4.8. Archival data was
extracted from the {\em Chandra} Webchaser and XMM data archives. 
The NASA/IPAC Extragalactic Database (NED), which is operated by 
JPL/Caltech, under contract with NASA was used throughout as were 
the ADS facilities and arXiv for the literature. 
We thank Gary Mamon, and Paul Nulsen for useful discussions of 
the Virgo cluster and gas stripping, respectively.  
This work was supported by {\em Chandra} grants 
GO1-13141X, GO1-12110X, GO0-1106X, NASA contract NAS8-03060, the 
University of Southampton and the Smithsonian Astrophysical Observatory.

\section*{ Appendix}
\label{appendix}

The {\em Chandra} data measure the X-ray emission from all sources in
M60; this will encompass the
diffuse galaxy gas, any unresolved point sources, such as
cataclysmic variables (CV's), accreting white dwarfs (AB's) and low
mass X-ray binaries (LMXBs), as well as the Virgo ICM emission along
the line of sight. We model the relative contribution of
each of these  unresolved stellar X-ray sources  
to the X-ray  luminosity of M60 below: 

\subsubsection*{Cataclysmic Variables and Accreting Binaries}
\label{sec:CVAB}

Based on X-ray observations of M32, Revnivtsev \etal (2007)
found that unresolved stellar objects may provide a large fraction of 
the diffuse
X-ray emission in low mass galaxies. In their follow-up paper,
Revnivtsev \etal (2008) show the old stellar populations in
galaxies can be characterized by a universal value of X-ray emissivity
per unit stellar mass or per unit K-band luminosity.

From studying the K-band images of M60 taken with the 2MASS survey, we
determined the stellar mass and implied X-ray luminosity of the
stars. The K-band luminosity measured in a $200\as$ radius 
region centered on the X-ray peak is 
$L_{\rm{K}} = 2.90\times10^{11}\,L_{{\rm K},\odot}$.
Bell \etal (2003)  showed that the relation between
the galaxy's stellar mass and the K-band luminosity can be 
expressed as:

\begin{equation}
	\log\left(\frac{M_{*}}{L_{\rm{K}}}\right) = a_{\rm{K}} +
        b_{\rm{K}}\times(B-V)
\label{eq:Kband}
\end{equation}
where $M_*$ is in units of $\Ms$ and $L_{\rm{K}}$ is in units of 
$L_{\rm{K},\odot}$. 
Using $a_{\rm{K}}=-0.206$, $b_{\rm{K}}=0.135$ (Bell \etal 2003) and 
the extinction corrected color for M60 $B-V=0.95$ from RC3 data 
(de Vaucouleurs \etal 1991; NED) in Eq. \ref{eq:Kband}, we find a 
stellar mass of $M_{*}=2.4\times10^{11} M_{\odot}$, where systematic
uncertainties in the M/L relationship may be as high as $25\%$.  

Revnivtsev \etal (2007) give the relation between X-ray luminosity
and stellar mass in the soft ($0.5-2.0$\,keV) energy band as follows: 

\begin{equation}
	L_{\rm{X}}^{0.5-2.0 \rm{keV}} =
        7\times10^{38}\left(\frac{M_{*}}{10^{11}M_{\odot}}\right)
        \rm{erg~ s}^{-1}
\label{eq:starmass}
\end{equation}

Using Eq. \ref{eq:starmass}, the stellar contribution to the soft
X-rays is therefore
$L_{\rm{X}}^{0.5-2.0 \rm{keV}}= 1.68\times10^{39}$\ergs, such
that the stellar component is $1.5\%$  
of the total $0.5-2$\,keV emission. 
One should note, however, that there is of order a factor $2$ 
scatter in the
L$_{\rm X}$ - L$_{\rm K}$ relation for these components in early type
galaxies (see, e.g. Bogdan and Gilfanov 2011).

In fitting the X-ray
spectra, this component is modeled with a mekal model plus  power law 
with fixed temperature $kT=0.5$ keV, Anders and Grevasse abundance
$A=1.0\Zs$ and power law exponent $\Gamma=1.9$ (Revnivtsev \etal
2008). The relative normalization of the two components is fixed by
setting the ratio of mekal model to power law fluxes to be $2.03$
in the $0.5-2$\,keV band and the overall normalization  is fixed such
that the contribution of CV and AB stars is $1.5\%$ of the $0.5-2$\,keV flux. 

\subsubsection*{Low Mass X-ray Binaries}
\label{sec:LMXB}

A second component to the X-ray emission is required to
account for unresolved LMXBs below the individual source detection
threshold that contribute to the overall diffuse X-ray luminosity. 
The azimuthally averaged spatial
distribution of the number of LMXBs for most normal
galaxies follows closely the distribution of the near-infrared 
light (Gilfanov 2004). The combined luminosity functions of
LMXBs for such galaxies are as follows:

\begin{equation}
	\frac{\mathrm{d}N}{\mathrm{d}L_{38}} = \left\{
	\begin{array}{l l}
		K_{1} (\frac{L_{38}}{L_{\rm{b},1}})^{-\alpha_{1}}, & L_{38} < L_{\rm{b},1} \\
		K_{2} (\frac{L_{38}}{L_{\rm{b},2}})^{-\alpha_{2}}, & L_{\rm{b},1} < L_{38} < L_{\rm{b},2} \\
		K_{3} (\frac{L_{38}}{L_{\rm{cut}}})^{-\alpha_{3}}, & L_{\rm{b},2} < L_{38} < L_{\rm{cut}} \\
		0, & L_{38} > L_{\rm{cut}} 
	\end{array}
	\right.
\label{eq:lumfunc}
\end{equation}

\begin{equation}
\begin{array}{l l}
	K_{2}= & K_{1}(\frac{L_{\rm{b},1}}{L_{\rm{b},2}})^{\alpha_{2}}, \\
	K_{3}= & K_{2}(\frac{L_{\rm{b},2}}{L_{\rm{cut}}})^{\alpha_{3}},  \\
	L_{38}= &\frac{L_{\rm{X}}}{10^{38}} \rm{erg~ s}^{-1}
\end{array}
\label{eq:lumfundef}
\end{equation}

The average normalization is $K_{1}=440.4\pm25.9$ per $10^{11}
\Ms$, $\alpha_{1}=1.0$, $\alpha_{2}=1.64$, $L_{\rm{b},1}=0.19$,
$L_{\rm{b},2}=5.1$ (Gilfanov 2004).  The total number of
LMXBs and their collective luminosity is directly proportional to the
stellar mass of the host galaxy. In regions without significant
diffuse X-ray emission, $7$ counts is typically sufficient to detect a 
point source with Chandra. Assuming a power law LMXB spectrum with a
slope of $\Gamma = 1.56$, and the combined exposure time of 
$262$\,ks, the estimated source detection sensitivity would be $6 \times
10^{36}$\ergs. However, due to the presence of copious diffuse
emission in M60, the actual source detection sensitivity is
significantly higher than this. Based on a sensitivity map that was 
computed using the CIAO \textsc{lim\_sens} task, we estimate that in
most regions of M60 we detect sources brighter than 
$L_{\rm lim} = 2 \times 10^{37}$\ergs.  Based on this source detection 
sensitivity and equations \ref{eq:lumfunc} 
and \ref{eq:lumfundef},we predict that the X-ray luminosity of unresolved 
LMXBs in M60 is
$L_{\rm{LMXB}}^{200\as} = 2.6 \times 10^{39}$\ergs in the $0.5-2$\,keV
band, approximately $2\%$ of 
of the total X-ray luminosity within the $200\as$ radius region.

\end{document}